\documentclass[aps,prl,twocolumn,10pt,superscriptaddress,nofootinbib,balancelastpage]{revtex4-1} 
\usepackage[latin1]{inputenc}
\usepackage{amsmath,amssymb}
\usepackage{mathrsfs} 
\usepackage[capitalise]{cleveref}
\usepackage{calc}
\usepackage{dsfont}
\usepackage{color}
\usepackage{seqsplit}
\usepackage{ifthen}
\usepackage{graphicx}

\usepackage{mathtools} %Fuer Indizes oben links

\allowdisplaybreaks

\newcommand{\ii}{\mathrm{i}}%
\newcommand{\dif}{\mathrm{d}}%
\newcommand{\Tr}{\operatorname{Tr}}%
\newcommand{\ZT}[1]{\textquotedblleft#1\textquotedblright}%

\newlength{\myl}%
\newcommand{\SUM}[2]{{\setlength{\myl}{\widthof{$\displaystyle\sum_{#1}^{#2}$}*\real{0.5}-\widthof{$\displaystyle\sum$}*\real{0.5}}\sum_{#1}^{#2}\;\hspace{-\the\myl}}}% Summen in abgesetzten Gleichungen
\newcommand{\INT}[3]{\settowidth{\myl}{$\displaystyle\int_{#1}^{#2}$}{\int_{#1}^{#2}\;\;\;\hspace{-\the\myl}\dif #3}\,}% Integrale in abgesetzten Gleichungen
\newcommand{\TINT}[3]{\settowidth{\myl}{$\int_{#1}^{#2}$}{\int_{#1}^{#2}\!\ifthenelse{\equal{#1#2}{}}{}{\;\;\;\;\hspace{-\the\myl}}\dif #3}\,}%
\newcommand{\EINT}[3]{\settowidth{\myl}{$\int_{#1}^{#2}$}{\int_{#1}^{#2}\;\;\;\,\hspace{-\the\myl}\dif #3}\,}% Integrale in Exponenten

\begin{document}
\title{Mori-Zwanzig formalism for general relativity:\\a new approach to the averaging problem}

\author{Michael te Vrugt}
\affiliation{Institut f\"ur Theoretische Physik, Center for Soft Nanoscience, Westf\"alische Wilhelms-Universit\"at M\"unster, D-48149 M\"unster, Germany}

\author{Sabine Hossenfelder}
\affiliation{Frankfurt Institute for Advanced Studies, D-60438 Frankfurt am Main, Germany}

\author{Raphael Wittkowski}
\email[Corresponding author: ]{raphael.wittkowski@uni-muenster.de}
\affiliation{Institut f\"ur Theoretische Physik, Center for Soft Nanoscience, Westf\"alische Wilhelms-Universit\"at M\"unster, D-48149 M\"unster, Germany}

\begin{abstract}
Cosmology relies on a coarse-grained description of the universe, assumed to be valid on large length scales. However, the nonlinearity of general relativity makes coarse-graining extremely difficult. We here address this problem by extending the Mori-Zwanzig projection operator formalism, a highly successful coarse-graining method from statistical mechanics, towards general relativity. Using the Buchert equations, we derive a new dynamic equation for the Hubble parameter which captures the effects of averaging through a memory function. This gives an empirical prediction for the cosmic jerk.
\end{abstract}
\maketitle

General relativity \cite{Einstein1914,Einstein1915b,Einstein1916} is a highly successful theory which has been used to model the universe on many different scales. On the \ZT{microscopic} scale of individual stars and stellar black holes, it has been tested and confirmed to great accuracy by a large number of experiments, ranging from the gravitational deflection of light by the sun \cite{DysonED1920} to the recent direct detection of gravitational waves by the LIGO and Virgo collaborations \cite{Abbott2016,Abbott2016b}. However, we also use general relativity on \ZT{macroscopic} scales in cosmology, where we interpret it as a coarse-grained description of the universe on the average. 

This methodology is mathematically suspect. Einstein's field equations are highly nonlinear, therefore they cannot hold for averaged quantities the same way as for the non-averaged ones. This means if those equations are valid on the scale of individual stars, then they should have correction terms on cosmological scales \cite{ClarksonELU2011}. Finding a coarse-grained description for the lumpy universe is often referred to as the \ZT{averaging problem} \cite{ClarksonELU2011,Wiltshire2007,Wiltshire2011,Zalaletdinov2008,Ginat2021,Paranjape2009,Buchert2011}, and the effect of inhomogeneities on averaged quantities is known as \ZT{backreaction} \cite{ClarksonELU2011,Rasanen2011}.
 
Some authors have suggested that backreaction can mimic effects of (and can therefore constitute an alternative to) dark energy, but the quantitative importance of such effects has remained controversial \cite{Kolb2011,ClarksonELU2011,BuchertEtAl2015,Buchert2005}. Understanding how to properly average Einstein's field equations has only become more pressing with recent observations that indicate a tension between different measurements of the Hubble parameter (\ZT{Hubble tension}) \cite{PoulinSKK2019,JedamzikP2020,BerghausK2020,Freedman2017,Riess2020}. A large variety of solutions for this issue have been discussed \cite{DiValentinoEtAl2021}, including an explanation based on cosmic inhomogeneities \cite{HeinesenB2020,Bolejko2018,MacphersonLP2018}.

Among the most popular theoretical approaches to the averaging problem are the \textit{Buchert equations} \cite{Buchert2000,Buchert2001,BuchertMR2020}, which provide a simple model for the time evolution of averaged scalar quantities. With them, an inhomogeneous universe can be described with an effective (nonequilibrium) equation of state \cite{Buchert2005}. Coarse-grained descriptions of nonequilibrium systems are of importance also in other fields of physics. There, systematic coarse-graining techniques have been developed, of which the \textit{Mori-Zwanzig projection operator formalism} \cite{Mori1965,Zwanzig1960,Nakajima1958,teVrugtW2019,MeyerVS2017,MeyerVS2019} is perhaps the most powerful one. 

This formalism, which is the focus of this work, employs a projection operator to reduce the complete microscopic dynamics to that of a subset of a priori arbitrary  \ZT{relevant variables}. The remaining \ZT{irrelevant variables} become encoded in a generalized Langevin equation as memory and noise terms. (For an introduction, see Refs.\ \cite{Grabert1982,teVrugtW2019d,KlipensteinTJSvdV2021,RauM1996,Schilling2021}.) Applications of the Mori-Zwanzig formalism include the derivation of dynamical density functional theory \cite{teVrugtLW2020,Yoshimori2005,EspanolL2009,WittkowskiLB2012,WittkowskiLB2013,WittkowskitVJLB2021}, hydrodynamics \cite{WittkowskitVJLB2021,CamargodlTDZEDBC2018}, glassy systems \cite{Das2004}, philosophy of physics \cite{teVrugt2020,Robertson2018,Wallace2015}, solid-state theory \cite{Fulde1995,KakehashiF2004}, and spin relaxation theory \cite{teVrugtW2019,KivelsonO1974,Bouchard2007}. The formalism has also been successfully applied in high-energy physics and relativistic causal hydrodynamics \cite{HuangKKR2011,KoideK2008,KoideNK2009,Koide2007,KoideM2000,KoideMT1999}. Below, we will extend the Mori-Zwanzig formalism to general relativity. 
 
We start by introducing the Mori-Zwanzig formalism following Ref.\ \cite{teVrugtW2019d}. The microscopic equation of motion for an arbitrary variable $A$ obeying Hamiltonian dynamics can be written as $\dot{A} = \ii L A$, where the overdot denotes a derivative with respect to time $t$, and $L$ is the Liouvillian defined as the Poisson bracket (or, in the quantum case, the commutator) with the Hamiltonian of the respective theory. We can  formally integrate this equation to $A(t) = e^{\ii L t} A$ (assuming that $L$ is time-independent, otherwise one needs to time-order the exponent \cite{teVrugtW2019}). Here, $A$ is the (time-independent) Schr\"odinger-picture observable and $A(t)$ is the (time-dependent) Heisenberg-picture observable. At the reference time $t=0$, Schr\"odinger and Heisenberg-picture observables agree \cite{teVrugtW2019d,BalianV1985}. Although the distinction between Schr\"odinger picture and Heisenberg picture is known mostly from quantum mechanics, it can be used also in the classical case \cite{HolianE1985}.

We then choose an arbitrary set of relevant variables $\{A_k\}$ and define the projection operator acting on a phase-space function $X$ as
\begin{equation}
PX = A_j(A_j,A_k)^{-1}(X,A_k),  
\end{equation}
where $(X,Y)$ is a scalar product, often defined as $(X,Y) = \Tr(\hat{\rho}XY)$ with the trace $\Tr$ (denoting an integral over phase space in the classical and a quantum-mechanical trace in the quantum case) and a probability distribution $\hat{\rho}$. Indices appearing twice are summed over. 

Next, we apply the Dyson identity
\begin{equation}
e^{\ii Lt} = e^{\ii QLt} + \INT{0}{t}{s}e^{\ii L(t-s)} P\ii Le^{\ii Q Ls}
\label{identity}%
\end{equation}
with the orthogonal projection operator $Q = 1-P$ to the expression $Q\ii L A_i$ which gives us the \textit{Mori-Zwanzig equation} 
\begin{equation}
\dot{A}_i(t) = \Omega_{ij} A_j(t) + \INT{0}{t}{s}K_{ij}(s)A_j(t-s) + F_i(t)
\label{exact}
\end{equation}
with the frequency matrix $\Omega_{ij} = (A_j,A_k)^{-1}(\ii L A_i,A_k)$, the memory matrix $K_{ij}(t)=(A_j,A_k)^{-1}(\ii L F_i(t),A_k)$, and the random force $F_{i}(t)=e^{\ii QLt} Q \ii LA_{i}$. Equation \eqref{exact} is an exact transport equation for the relevant variable $A_i$. We set the lower bound of the time integral in \cref{exact} to $t=0$, but different values are possible \cite{Grabert1978,Grabert1982}. 

We now apply the Mori-Zwanzig formalism to general relativity. In its standard form, the Mori-Zwanzig formalism is based on solving an initial value problem $\dot{A}= \ii L A$ with some Liouvillian $L$ in the form $A(t)=\exp(\ii L t)A(0)$. Thus, to apply the Mori-Zwanzig formalism, we first write the equations of general relativity as a (Hamiltonian) initial value problem using the \textit{Arnowitt-Deser-Misner (ADM) }formalism \cite{ArnowittDM1959,Arnowitt1DM960,ArnowittDM1960b}, reviewed in Refs.\ \cite{SchaeferJ2018,Lehner2001}. 

In the {ADM} formalism, one foliates spacetime into spacelike hypersurfaces $\Sigma_t$, labelled by a time-coordinate $t$. The canonical variables (in addition to those for matter) are the spatial metric on the hypersurfaces $\gamma_{ij}$ and its conjugate momentum $\pi^{ij}$, which satisfy the dynamic equations $\dot{\gamma}_{ij} =\ii L \gamma_{ij}=\delta \mathcal{H}/\delta \pi^{ij}$ and $\dot{\pi}^{ij}=\ii L \pi^{ij}=-\delta \mathcal{H}/\delta \gamma_{ij}$ (where $\ii L$ denotes the Poisson bracket with the Hamiltonian $\mathcal{H}$), as well as the functions $N^i$ (shift) and $N$ (lapse) that can be chosen freely \cite{Gourgoulhon2007}. Hence, the derivation of the Mori-Zwanzig equation \eqref{exact} can be used also in general relativity. 

However, the {ADM} formalism contains also the non-dynamical \textit{constraint equations} $\delta \mathcal{H}/\delta N^i=0$ and $\delta \mathcal{H}/\delta N = 0$ \cite{Gourgoulhon2007}. These have to be satisfied by the initial conditions in order for the solution of the initial value problem to be a solution of Einstein's field equations; their validity is then preserved by the time evolution \cite{Gourgoulhon2007,Lehner2001}. In the Mori-Zwanzig formalism, the constraints have to be taken into account not only in the initial conditions, but also in the definition of the projection operator. If a constraint equation requires that $A=B$ with two observables $A$ and $B$, $P$ has to be defined in such a way that $PA=PB$ holds (otherwise, one might group a part of the dynamics to the random force $F_i$ that actually belongs to the organized drift $\Omega_{ij}A_j$). This can be ensured by choosing $\hat{\rho}$ so that it assigns zero probability to configurations that do not satisfy the constraint equations. Apart from this, $\hat{\rho}$ can be adapted to the problem at hand. For the problem we are interested in, a good choice for $\hat{\rho}$ is a distribution that assigns equal probability to all configurations compatible with the observed present value of the Hubble parameter $H_0$. Moreover, in general relativity, the phase-space integral in the definition of the  scalar  product becomes, for $\gamma_{ij}$ and $\pi^{ij}$, a functional integral. Finally, the lower limit of the time integral in \cref{exact} typically corresponds to the time of the preparation of the system, although later times can also be used \cite{Grabert1978}. In cosmological contexts, the \ZT{time of preparation} would be the big bang, which can be a problematic choice of the reference time due to singularities. Hence, a slightly later time might be more appropriate. An earlier time should not be chosen, since otherwise the system would have to \ZT{remember} the values of the observables from times before the big bang. 

We now demonstrate the usefulness of the Mori-Zwanzig formalism in general relativity by applying it to the cosmological averaging problem. Averaged scalar quantities in general relativity can be described using the Buchert equations \cite{Buchert2000,Buchert2001,BuchertMR2020}. For simplicity, we consider a universe filled with irrotational dust as done in Ref.\ \cite{Buchert2000}, but since Buchert's formalism can also be applied to more general cases \cite{BuchertMR2020}, our approach is not restricted to this simple example.

Let $D \subset \Sigma_t$ be the domain of the hypersurface $\Sigma_t$ we average over \cite{Behrend2008}. This domain has the volume $V$. We introduce the effective scale factor $a=(V/V_0)^\frac{1}{3}$, where $V_0$ is the volume at the present time. Our relevant variables are the effective Hubble rate $H = \dot{a}/a$, the squared effective Hubble rate $H^2$, and the cosmological constant $\Lambda$. The latter two are included since the contributions from $H^2$ and $\Lambda$ would otherwise be hidden in the memory kernel because \cref{exact} is linear in the relevant variables. Our relevant variables have to be functions of the phase-space variables. This is the case here since $H$ can be written as $H = -(\TINT{\Sigma_{t}}{}{^3r}\pi)/(6\TINT{\Sigma_{t}}{}{^3r}\sqrt{\gamma})$, where $\vec{r}$ is the position on the hypersurface $\Sigma_t$, $\pi = \gamma_{ij}\pi^{ij}$ the trace of the momentum $\pi^{ij}$, and $\gamma$ the determinant of the spatial metric $\gamma_{ij}$. The expression for $H$ follows from Buchert's result $H=-(\TINT{\Sigma_t}{}{^3r}\sqrt{\gamma}\mathcal{K})/(3\TINT{\Sigma_t}{}{^3r}\sqrt{\gamma})$ \cite{Buchert2000} together with $\pi^{ij} = \sqrt{\gamma} (\mathcal{K}\gamma^{ij} - \mathcal{K}^{ij})$ \cite{Gourgoulhon2007}, where $\mathcal{K}^{ij}$ is the extrinsic curvature tensor and $\mathcal{K}=\gamma_{ij}\mathcal{K}^{ij}$ its trace. 

As shown in Ref.\ \cite{Buchert2000}, the 3+1-decomposition with $N^i=0$ and $N=1$ gives the Buchert equations
\begin{align}
3(\dot{H}+H^2) + 4\pi G\bar{\rho} - \Lambda &= Q_D,\label{buchert1}\\
3H^2 - 8\pi G \bar{\rho}+ \frac{1}{2}\bar{R}_D- \Lambda &= - \frac{Q_D}{2}\label{buchert2},
\end{align}
where $G$ is Newton's gravitational constant, $\bar{\rho}$ the averaged mass density (due to mass conservation given by $\bar{\rho} = M/(V_0 a^3)$ with the total mass $M$), $\bar{R}_D$ the averaged spatial Ricci scalar, and $Q_D$ the backreaction term. Equation \eqref{buchert1} is the averaged Raychaudhuri equation and \cref{buchert2} is an averaged Hamiltonian constraint. $Q_D$ and $\bar{R}_D$ are related via the integrability constraint $\partial_t (a^6 Q_D) + a^4\partial_t(a^2 \bar{R}_D)=0$. The Buchert equations \eqref{buchert1} and \eqref{buchert2} are similar in form to the Friedmann equations \cite{Friedman1922,Friedmann1924}, which are the standard description of an isotropic homogeneous universe, and obtain the same form (for a flat universe) when setting $\bar{R}_D = Q_D = 0$. We employ units in which the speed of light $c$ is 1. 

Equation \eqref{buchert1} is the microscopic equation of motion for $H$. Next, we eliminate $\bar{\rho}$ by solving \cref{buchert2} for $\bar{\rho}$ and inserting the result into \cref{buchert1}, which gives
\begin{equation}
\dot{H} = -\frac{3}{2}H^2+ \frac{1}{2}\Lambda + \mathfrak{F}
\label{microscopic}
\end{equation}
with $\mathfrak{F}=Q_D/4 - \bar{R}_D/12$. Compared to \cref{buchert1}, \cref{microscopic} has the advantage that the projection $P$ (which has to respect the constraint \eqref{buchert2}) is easier to apply. Applying $\exp(\ii L t$) to \cref{microscopic} and using \cref{identity} gives
\begin{align}
\dot{H}(t)&= -\frac{3}{2}H^2(t)+ \tilde{\Omega}_{HH}H(t) + \tilde{\Omega}_{HH^2} H^2(t) + \frac{1}{2}\tilde{\Lambda}(t) \notag\\
&\quad\, +\INT{0}{t}{s} \big(K_{HH}(s)H(t-s) + K_{HH^2}(s)H^2(t-s)\big) \notag\\
&\quad\, +F_H(t),
\label{buchert1projected}
\end{align}
where we have introduced the frequencies 
\begin{equation}
\tilde{\Omega}_{H A_j} = (A_j,A_k)^{-1}(\mathfrak{F},A_k)
\end{equation}
with $A_j\in\{H,H^2,\Lambda\}$, the shifted cosmological constant
\begin{equation}
\tilde{\Lambda}(t)= \Lambda\bigg(1 + 2\tilde{\Omega}_{H\Lambda} + 2\INT{0}{t}{s}K_{H\Lambda}(s)\bigg),    \label{shifted}
\end{equation}
the memory functions
\begin{align}
K_{H A_j}(t)=(A_j,A_k)^{-1}(\ii L F_H(t),A_k),\label{memory1}  \end{align}
and the random force (noise) 
\begin{equation}
F_{H}(t)=e^{\ii QLt}Q\mathfrak{F}.\label{randomforcebuchert}
\end{equation}
The noise $F_H$ is related to the memory kernels via \cref{memory1}. These equations constitute a cosmological fluctuation-dissipation theorem (FDT), since they are analogous to the usual second FDT \cite{Grabert1978,KlipensteinTJSvdV2021}. Equation \eqref{buchert1projected} is a special case of the general result \eqref{exact}. The tilde over the frequencies emphasizes that we have split off the terms appearing also in \cref{microscopic} to make them visible explicitly.

At first glance, it might appear as if \cref{buchert1projected} is no improvement compared to \cref{microscopic}, since it contains significantly more terms. However, this impression is misleading. Equation \eqref{microscopic} contains the function $\mathfrak{F}$ about which (without further input) we know next to nothing. It could also depend on $H$. In \cref{buchert1projected} however, all contributions from the relevant variables are made explicit. The additional terms are simple and linear in the relevant variables, and the random force $F_H$ really only contains the \ZT{irrelevant} part of the dynamics, such that (with some caveats, see Ref.\ \cite{EspanolO1993}) it can be thought of as a noise term. (It can still depend on $a$, but even this dependence could be extracted by including $a$ in the set of relevant variables. This is not done here since our main purpose is to explain the formalism.) Equations of the form \eqref{buchert1projected} are well understood, and this understanding can now be exploited to understand better the effects of backreaction. In particular, there are well-established numerical methods for extracting the form of the memory functions from measured data of the observable of interest \cite{MeyerPS2020,MeyerWSS2020,BerkowitzMKM1981}.

To solve \cref{buchert1projected}, we assume that the noise is small and can be omitted. (Alternatively, if we gave the random force a stochastic interpretation, \cref{buchert1projected} could be interpreted as giving rise to a probability distribution over possible universes, corresponding to different noise functions.) Next, we make a simple ansatz for the memory kernels of the form $K_{HH^i}(t)= - \kappa_i \exp(-\xi_i t)$ with memory amplitude $\kappa_i$ and damping coefficient $\xi_i$, where $i=1,2$. Such an ansatz is common in high-energy physics \cite{KoideKR2006,KoideDMK2007}. It has the advantage that it allows to remove the convolution integral in \cref{buchert1projected} (see Refs.\ \cite{KoideKR2006,teVrugtJW2021}). 

Finally, we make the, also common, assumption that $K_{H\Lambda}$ relaxes quickly (Markovian approximation \cite{RauM1996}), such that the third term on the right-hand-side of \cref{shifted} is approximately constant. The backreaction then just changes the measured value of the cosmological constant compared to the \ZT{micoscopic} case. In other applications of the Mori-Zwanzig formalism, the irrelevant dynamics typically contributes mainly to the linear memory term, such that we set $\tilde{\Omega}_{HH}=0$, $\tilde{\Omega}_{HH^2}=0$, and $\kappa_2=0$ to focus on the dominant contribution. Then, one can rewrite \cref{buchert1projected} as
\begin{equation}
\ddot{H}=-3H\dot{H} -\xi_1\bigg(\dot{H}+\frac{3}{2}H^2-\frac{1}{2}\tilde{\Lambda}\bigg)- \kappa_1 H, \label{eins}
\end{equation}
where we have suppressed the time-dependence.

We now link our results to astronomical parameters, in particular to the redshift $z$ and the jerk $j$. If structures evolve sufficiently slowly, $z$ is determined by the averaged quantity $a$ through the relation $a=1/(1+z)$ \cite{HeinesenB2020}. It follows from the definition of $H$ that 
\begin{equation}
\ddot{H} = \frac{\dddot{a}}{a} -3 H \dot{H} - H^3.
\label{zwei}
\end{equation}
Using the definition of the jerk \cite{MukherjeeB2016}
\begin{equation}
j=-\frac{\dddot{a}}{aH^3}=-1 +(1+z)\frac{(H^2)'}{H^2}-\frac{1}{2}(1+z)^2\frac{(H^2)''}{H^2}, 
\label{jerk1}
\end{equation}
where $'$ denotes a derivative with respect to $z$ (from now on, we write $H=H(z)$), we find from \cref{eins,zwei,jerk1} 
\begin{equation}
j=-\frac{-\xi_1 (\dot{H}+ \frac{3}{2}H^2 - \frac{1}{2}\tilde{\Lambda})- \kappa_1 H + H^3}{H^3}.
\label{jerk2}
\end{equation}
(Note that some authors define $j=+\dddot{a}/(aH^3)$ instead \cite{Visser2004}.) With the plausible assumption that the universe behaves approximately Friedmannian on large scales, such that $\dot{H}\approx - \frac{3}{2}H^2 + \frac{1}{2}\tilde{\Lambda}$, \cref{jerk2} simplifies to
\begin{equation}
j= -1 + \frac{\kappa_1}{H^2}.
\label{jerk3} 
\end{equation}

In the widely used $\Lambda${\sc CDM} model, the expansion of the universe is dominated by the cosmological constant ($\Lambda$) and by cold dark matter ({\sc CDM}), and the jerk is $j=-1$ \cite{MukherjeeB2016}. For this reason, measurements of $j$ have been used to investigate possible deviations from the $\Lambda$CDM model \cite{LussoEtAl2019,ZhaiZZLZ2013,MukherjeeB2021}, which were suggested by various recent studies \cite{LussoEtAl2019,RisalitiL2019}. 
Deviations of $j$ from $-1$ have been linked to a non-standard equation of state for dark energy \cite{MukherjeeB2016} or energy transfer from matter to dark energy \cite{PerezSW2021}. Effects of backreaction on the jerk were also considered \cite{CarvalhoB2016,Bochner2013,Bochner2012}. 

The functional form of Eq.\ \eqref{jerk3} -- derived here from first principles -- has been proposed in Ref.\ \cite{MukherjeeB2016} as a possible ad-hoc modification of $\Lambda${\sc CDM}. There, it was shown that the differential equation obtained by setting \cref{jerk3} equal to \cref{jerk1} has the solution $H^2/H_0^2 = c_1(1+z)^3+1-c_1 +\frac{2}{3} j_1 \ln(1+z)$ with an integration constant $c_1$ and $\kappa_1 = j_1 H_0^2$. This gives us 
\begin{equation}
\bigg(\frac{\dot{a}}{a}\bigg)^2 = \frac{H_0^2 c_1}{a^3} + H_0^2(1-c_1) - \frac{2}{3}\kappa_1\ln(a).
\label{differentialequation}%
\end{equation}
Essentially, we have converted a differential equation for $H(t)$ into a (analytically solvable) differential equation for $H(z)$ whose solution (\cref{differentialequation}) is a differential equation for $a(t)$.  The first two terms on the right-hand side of \cref{differentialequation} will also be present in the $\Lambda${CDM} model. The third term, on the other hand, is solely a consequence of backreaction. 

In a homogeneous and isotropic universe, the Hamiltonian constraint $3H^2=8\pi G \rho + \Lambda$ (with mass density $\rho$) implies that $c_1 = 8\pi G \rho_0/(3 H_0^2)$ \cite{MukherjeeB2016}, where $\rho_0$ is the present value of $\rho$, and $H_0^2(1-c_1) = \Lambda/3$. (These relations follow directly from comparing the constraint with \cref{differentialequation} using $\rho = \rho_0/a^3$, since $\kappa_1 = 0$ in the homogeneous case.) However, in the coarse-grained description considered here, we have to use the averaged constraint \eqref{buchert2}. Hence, we cannot simply read off $c_1$, since it is possible that the backreaction also has a contribution $\propto a^{-3}$ (suggested in Ref.\ \cite{Buchert2005}). We thus proceed by treating $c_1$ as a constant that needs to be determined by measurement. 

Based on fits to various data sets (supernova distance modulus data, observational Hubble data, baryon acoustic oscillation data, and cosmic microwave background shift parameter data), \citet{MukherjeeB2016} have found the values $c_1 = 0.298 \pm 0.010$ and $j_1 = 0.078 \pm 0.140$. Hence, while the data is consistent with the standard result $j_1 = 0$ within a $1\sigma$ confidence region, the best fit value is $j_1 \neq 0$.  

\begin{figure}
\centering
\includegraphics[width=\linewidth]{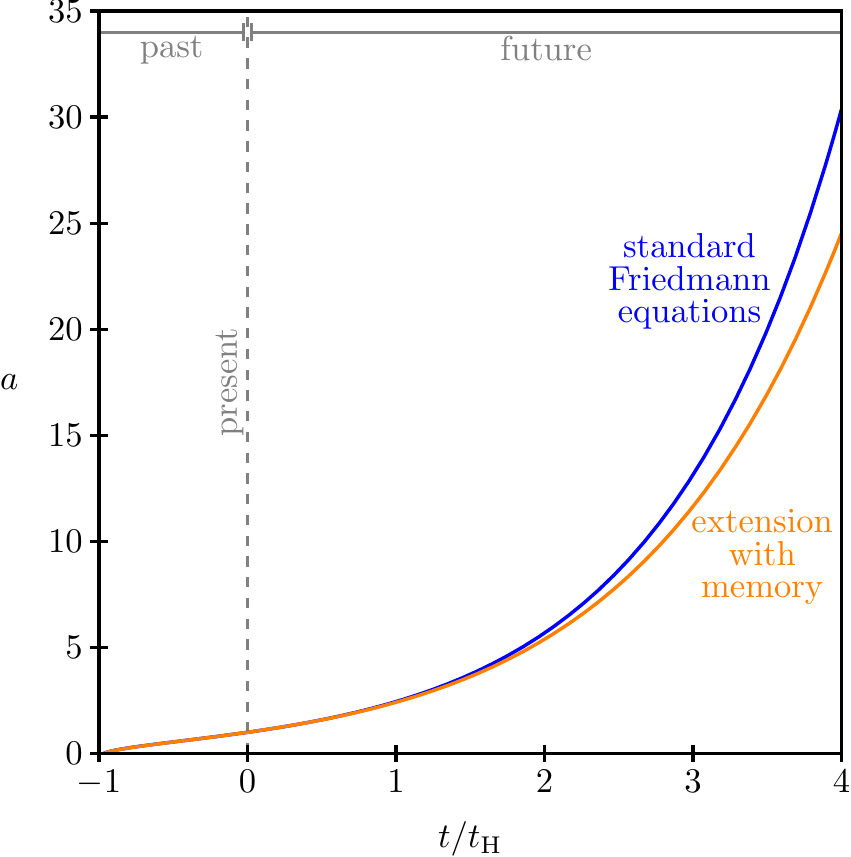}
\caption{\label{fig:1}Time evolution of scale factor $a(t)$, obtained by solving \cref{differentialequation}, for $\kappa_1/H_0^2 = 0$ (standard Friedmann equations) and $\kappa_1/H_0^2 = 0.078$ (extension with memory). We show only the expanding solution. Since \cref{differentialequation} is quadratic in $\dot{a}$, there exists a second solution that corresponds to a shrinking universe.}
\end{figure}

We have used this best-fit value to numerically integrate \cref{differentialequation}. The result is shown in \cref{fig:1} in comparison to the standard case without memory effects. In this figure, time is measured in units of the Hubble time $t_H = 1/H_0$ and $t=0$ corresponds (here) to the present. We use $\kappa_1/H_0^2 =0$ (standard Friedmann equations) and $\kappa_1/H_0^2 = 0.078$ (extension with memory), and set $c_1 = 0.298$ in both cases. As one sees, the backreaction-induced memory has the effect of slowing down the expansion of the universe.  

In conclusion, we have extended the Mori-Zwanzig formalism to general relativity and demonstrated its usefulness by deriving an expression for the backreaction in general relativity from first principles. We have shown that this approach predicts a relation between the cosmological jerk and the Hubble rate. Given the relevance of the Mori-Zwanzig formalism in other fields, we hope that our work provides a useful starting point for further investigations in precision cosmology and other areas of astrophysics. 

\acknowledgments
M.t.V.\ thanks Saul Perlmutter and Adam G.\ Riess for helpful discussions during the Lindau Nobel Laureate Meeting 2021 and the Studienstiftung des deutschen Volkes for financial support. 
S.H.\ and R.W.\ are funded by the Deutsche Forschungsgemeinschaft (DFG, German Research Foundation) -- grant numbers HO 2601/8-1 and WI 4170/3-1.

\nocite{apsrev41Control}
\bibliographystyle{apsrev4-1}
\bibliography{refs,control}
\end{document}